\documentclass[iop,revtex4]{emulateapj}

\usepackage{longtable}

\usepackage{natbib}

\usepackage{lscape}
\usepackage{amsmath}
\usepackage{color}
\usepackage{url}
\usepackage{ulem}
\usepackage{multirow}

\bibpunct{(}{)}{;}{a}{}{,}

\def\apj{ApJ}%
\def\apjl{ApJ}%
\def\apjs{ApJS}%
\def\ao{Appl.~Opt.}%
\def\apss{Ap\&SS}%
\def\aap{A\&A}%
\def\mnras{MNRAS}%
\def\ssr{Space~Sci.~Rev.}%
\slugcomment{To appear in The Astrophysical Journal}

\shorttitle{Compton-thick Accretion in the local Universe}
\shortauthors{Ricci et al.}

\begin{document}

\title{Compton-thick Accretion in the local Universe}

\author{C. Ricci\altaffilmark{1,2,3*}, Y. Ueda\altaffilmark{3}, M. J. Koss\altaffilmark{4,5}, B. Trakhtenbrot\altaffilmark{4,6}, F. E. Bauer\altaffilmark{1,2}, P. Gandhi\altaffilmark{7}}

\altaffiltext{1}{Pontificia Universidad Catolica de Chile, Instituto de Astrof'sica, Casilla 306, Santiago 22, Chile}
\altaffiltext{2}{EMBIGGEN Anillo, Concepcion, Chile}
\altaffiltext{3}{Department of Astronomy, Kyoto University, Oiwake-cho, Sakyo-ku, Kyoto 606-8502, Japan.}
\altaffiltext{4}{Institute for Astronomy, Department of Physics, ETH Zurich, Wolfgang-Pauli-Strasse 27, CH-8093 Zurich, Switzerland}
\altaffiltext{5}{Ambizione fellow}
\altaffiltext{6}{Zwicky fellow}
\altaffiltext{7}{School of Physics \& Astronomy, University of Southampton, Highfield, Southampton, SO17 1BJ}

\altaffiltext{*}{cricci@astro.puc.cl}

\begin{abstract}
Heavily obscured accretion is believed to represent an important stage in the growth of supermassive black holes, and to play an important role in shaping the observed spectrum of the Cosmic X-ray Background (CXB). Hard X-ray (E$>$10\,keV) selected samples are less affected by absorption than samples selected at lower energies, and are therefore one of the best ways to detect and identify Compton-thick (CT, $\log N_{\rm\,H}\geq 24$) Active Galactic Nuclei (AGN).
In this letter we present the first results of the largest broad-band (0.3--150\,keV) X-ray spectral study of hard X-ray selected AGN to date, focusing on the properties of heavily obscured sources. Our sample includes the 834 AGN (728 non-blazar, average redshift $z\simeq 0.055$) reported in the 70-months catalog of the all-sky hard X-ray {\it Swift}/BAT survey. We find 55 CT AGN, which represent $7.6^{+1.1}_{-2.1}\%$ of our non-blazar sample. Of these, 26 are reported as candidate CT AGN for the first time. We correct for  selection bias and derive the intrinsic column density distribution of AGN in the local Universe in two different luminosity ranges. We find a significant decrease in the fraction of obscured Compton-thin AGN for increasing luminosity, from $46\pm3\%$ (for $\log L_{\rm\,14-195} = 40-43.7$) to $39\pm3\%$ (for $\log L_{\rm\,14-195} = 43.7-46$). A similar trend is also found for CT AGN.
The intrinsic fraction of CT AGN with $\log N_{\rm\,H}=24-25$ normalised to unity in the $\log N_{\rm H} = 20-25$ range is $27\pm4\%$, and is consistent with the observed value obtained for AGN located within 20\,Mpc. 

\end{abstract}


\section{Introduction}

An obscured phase in the accretion history of supermassive black holes (SMBHs) is believed to represent a key step in the coevolution of Active Galactic Nuclei (AGN) and their host galaxies. This phase could be associated with a period of rapid SMBH accretion, and be part of an evolutionary scenario that starts with major galaxy mergers and ends with an unobscured SMBH (e.g., \citealp{Sanders:1988uq, Treister:2012kx}). 

Understanding the distribution of obscuration in AGN is also very important to fully comprehend the origin of the Cosmic X-ray background (CXB), which is due to the unresolved emission of AGN, and therefore is imprinted with the accretion history of the Universe. Synthesis models of the CXB (e.g., \citealp{Ueda:2003nx,Treister:2005ve,Ballantyne:2006kx,Gandhi:2007fk,Gilli:2007qf,Treister:2009uq,Akylas:2012uq,Ueda:2014fk,Comastri:2015cr}) have shown that a significant fraction of Compton-thick (CT, $\log N_{\rm H}\geq 24$) AGN are needed to reproduce the CXB. The intrinsic fraction of CT AGN ($f_{\rm CT}$) is however still highly uncertain, and spans from $\sim10\%$ to $\sim40\%$ of the total AGN population (e.g., \citealp{Brightman:2012ly}).

\begin{figure}[t!]
\centering
\includegraphics[width=9cm]{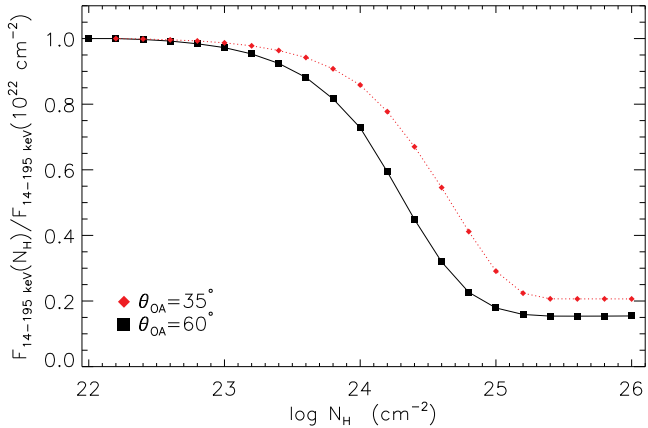}
  \caption{Observed 14-195 keV flux versus column density for the template spectral model of \citet{Ueda:2014fk}, assuming $\Gamma=1.8$ and an half-opening angle of the torus of $\theta_{\rm\,OA}=35^{\circ}$ and $\theta_{\rm\,OA}=60^{\circ}$. The fluxes are normalised to the value obtained for $\log N_{\rm\,H}=22$. The figure illustrates the attenuation curve expected for {\it Swift}/BAT.}
\label{fig:attenuation}
\end{figure}

Detecting and identifying CT AGN can be observationally challenging. At low redshift one of the best approaches is to use hard X-ray surveys ($E \gtrsim 10$\,keV), since the flux in this energy band is less affected by obscuring material than at lower energies, at least up to column densities of $N_{\rm H}\sim 10^{24}\rm\,cm^{-2}$ (Fig.\,\ref{fig:attenuation}). At high redshift it is possible to observe the rest-frame hard X-ray emission in the soft X-ray ($<10$\,keV) band, and recent deep surveys carried out with {\it Chandra} and {\it XMM-Newton} in the 0.3--10\,keV range (e.g., COSMOS, {\it Chandra} deep field south) have been able to find a considerable number of CT AGN at $z\gtrsim1$ (e.g., \citealp{Georgantopoulos:2013ly}, \citealp{Vignali:2014dq}, \citealp{Lanzuisi:2015zr}, \citealp{Buchner:2015ve}). In the local Universe hard X-ray detectors such as the Burst Alert Monitor (BAT, \citealp{Barthelmy:2005uq}) on board {\it Swift} \citep{Gehrels:2004kx}, IBIS/ISGRI on board {\it INTEGRAL} \citep{Winkler:2003fk} and FPMA/FPMB on board {\it NuSTAR} \citep{Harrison:2013zr} are therefore well suited to detect and classify CT AGN. 
{\it Swift}/BAT has been scanning the sky in the 14--195\,keV band since 2005, and given its complete coverage of the sky it can be used to infer the population characteristics of local heavily obscured AGN. However, previous works carried out both with {\it Swift}/BAT and {\it INTEGRAL} have reported only a handful of CT AGN (e.g., \citealp{Paltani:2008oq, Ajello:2008nx,Beckmann:2009fk,Malizia:2009oq,Burlon:2011uq,Ajello:2012fk,Vasudevan:2013ya}). Pointed {\it NuSTAR} observations have been very efficient in classifying and characterising CT AGN (e.g., \citealp{Balokovic:2014dq}, \citealp{Gandhi:2014bh}, \citealp{Arevalo:2014nx}, \citealp{Bauer:2014cr}, \citealp{Koss:2015qf}, \citealp{Lansbury:2015vn}, \citealp{Annuar:2015ve}, \citealp{Puccetti:2015qf}), although the sample of these heavily obscured AGN observed so far is still rather small.

Our group has recently carried out the largest study of broad-band X-ray emission of AGN (Ricci et al. in prep.), analysing in detail the 0.3--150\,keV spectra of the 834 AGN reported in the latest release (70-months, \citealp{Baumgartner:2013uq}) of the {\it Swift}/BAT catalog (Sect.\,\ref{Sect:specanalysis}). We present here the first results of our work, focused on the detection of 55 CT AGN, 26 of which are reported for the first time as candidate CT sources. We discuss our results in the framework of CXB synthesis models (Sect.\,\ref{sect:ctagn}), and use them to constrain the intrinsic $N_{\rm H}$ distribution and the intrinsic fraction of CT AGN (Sect.\,\ref{sect:intrinsicNH}).

\section{Sample and X-ray spectral analysis}\label{Sect:specanalysis}
Our total sample contains the 834 AGN reported in the 70-months {\it Swift}/BAT catalog \citep{Baumgartner:2013uq}, of which 106 are blazars. Blazars were identified based on the Rome BZCAT \citep{Massaro:2015ys} and on recent literature. In the following we will refer only to the 728 non-blazar AGN. The sample is local, with an average redshift $z\simeq 0.055$. We collected the best X-ray data below 10\,keV available as of March 2013, using spectra from {\it XMM-Newton}, {\it Chandra}, {\it Suzaku}, {\it Swift}/XRT and {\it ASCA}. We also examined the X-ray spectra of the $\sim 60$ objects reported as unknown in the {\it Swift}/BAT catalog (i.e. without a clear identification), and found that none of these objects show typical characteristics of CT AGN.
Obscured AGN were fitted with a model that includes: i) a primary X-ray emission source in the form of an absorbed cutoff power-law; ii) an unobscured reflection component (using a slab reflection model\footnote{\textsc{pexrav} in XSPEC \citep{Magdziarz:1995pi}}); iii) a scattered component in the form of a cutoff power-law, which parameters were fixed to those of the primary X-ray emission. This component was scaled by a constant ($f_{\rm\,scatt}$), which was left free to vary and had a typical value of a few percent; iv) a Gaussian to represent Fe K$\alpha$ emission line; v) emission from collissionally ionized plasma; vi) a cross-calibration constant to take into account possible flux variability between the soft X-ray observations and the 70-months averaged hard X-ray emission. Additional Gaussian lines were included to remove possible residuals in the iron region or below 4\,keV. In order to better constrain the column density and intrinsic flux, AGN with $N_{\rm\,H}$ consistent with $\geq 10^{24}\rm\,cm^{-2}$ within their 90\% uncertainties were then fitted with the physical torus model of \citet{Brightman:2011oq}, which considers absorption and reflection for a spherical-toroidal geometry. If statistically required, we added to this model a scattered component, additional Gaussian lines, a cross-calibration constant and collissionally ionized plasma.
Depending on the number of counts $\chi^2$ ($\sim 700$ AGN) or Cash ($\sim 130$ AGN) statistic were used to fit the X-ray spectra. Throughout this work we assume a cosmological model with $H_{0}=70\rm\,km\,s^{-1}\,Mpc^{-1}$, $\Omega_{\mathrm{m}}=0.3$ and $\Omega_{\Lambda}=0.7$. The spectral models used have been well tested in the past (e.g., \citealp{Gandhi:2014bh,Balokovic:2014dq,Lansbury:2015vn}) and follows our rich experience in dealing with X-ray spectra of AGN (e.g., \citealp{Ricci:2014dq,Bauer:2014cr,Gandhi:2014bh,Koss:2015qf}). Given the homogeneous approach used for the spectral modelling, our sample is extremely well suited for future multi-wavelength studies.

In Table\,\ref{tab:CTlist} we report the list of the 55 AGN identified as CT from our study (i.e., with a best-fit column density of $\log N_{\rm\,H}\geq 24$), the 26 sources classified here for the first time as candidate CT by means of X-ray spectroscopy are marked in boldface. The values of the parameters listed in the table were obtained by using the torus model of \citet{Brightman:2011oq}, as described above. The table also reports the values of the Fe K$\alpha$ EW and of the photon index ($\Gamma_{2-10}$) obtained by fitting the observed 2--10\,keV spectrum with a simple power law model. Both these parameters are diagnostics of heavy obscuration: the Fe K$\alpha$ EW is enhanced by the depletion of the X-ray primary emission due to the line-of-sight absorption, while the measured continuum in the 2--10\,keV band is much flatter than in unobscured AGN due to the larger influence of the reprocessed X-ray emission from distant material. For NGC\,1068 and NGC\,3079 we calculated $\Gamma_{2-10}$ in the 3--10\,keV range to reduce the strong influence of the radiative recombination continuum. All CT sources show a 2--10\,keV X-ray continuum significantly flatter than typical unobscured or Compton-thin AGN ($\Gamma\sim 1.8$, e.g. \citealp{Vasudevan:2013ya}). The table also reports the value of the photon index ($\Gamma$) obtained by our broad-band X-ray fit. The values of $N_{\rm\,H}$ we obtained for previously-known CT AGN are in agreement with previous works carried out by using physical torus models (e.g., \citealp{Gandhi:2014bh} and references therein, \citealp{Koss:2015qf,Balokovic:2014dq,Brightman:2015cr}). Details on the data reduction and fitting procedure, together with discussion on the individual CT sources,  will be reported in the X-ray catalogue (Ricci et al. in prep.).

\section{Compton-thick AGN and the Cosmic X-ray Background}\label{sect:ctagn}

Observational constraints on the fraction of CT sources are of fundamental importance to discern between different synthesis models of the CXB. With our study we found that 55 sources have best-fit column densities N$_{\rm H}\geq 10^{24}\rm\,cm^{-2}$, which represent $7.6\%$ of our hard X-ray selected sample. The number of CT AGN ranges from 40--63 (5.5--8.7\%) when 90\% confidence errors on the column density are considered. Therefore the fraction of CT AGN of our sample is $f_{\rm\,CT}=7.6^{+1.1}_{-2.1}\%$. This fraction would be at most $\sim 0.6$\% lower if all the objects reported as unknown in the {\it Swift}/BAT catalog were found to be non-blazar AGN.

The new CT AGN are located on average at higher redshifts compared to previously known hard X-ray selected CT sources ($z=0.042$ vs. $0.017$). A Kolomogorov-Smirnov (KS) test gives a p-value of $\sim0.2\%$ that the two samples are drawn from the same distribution. The newly detected CT AGN have on average higher luminosities ($\log L_{14-150}=44.23$ vs. $43.87$, KS p-value of $\sim10\%$) compared to previously known CT sources.  The average values of the photon indices are consistent between the two populations ($\Gamma=2.09\pm 0.05$ vs. $\Gamma=2.06\pm 0.05$, KS p-value of $\sim86\%$). While the average values of the column density of newly identified CT AGN ($\log N_{\rm H}=24.59$) is consistent with that of previously known CT AGN ($\log N_{\rm H}=24.63$), their distribution are significantly different (KS p-value of $\sim1\%$), with the distribution of new CT AGN peaking at lower values of the column density. In an independent search for Compton-thick AGN, Koss et al. (in prep.) selected objects from the {\it Swift}/BAT catalog via their spectral curvature at $E>14$\,keV, and follow-up {\it NuSTAR} observations have confirmed their CT nature. We have checked our sample and found that all of our suggested CT objects at $z\leq 0.03$ show similar high energy curvature, validating this technique for finding CT AGN.

\begin{figure}[t!]
\centering
\centering
\includegraphics[width=9cm]{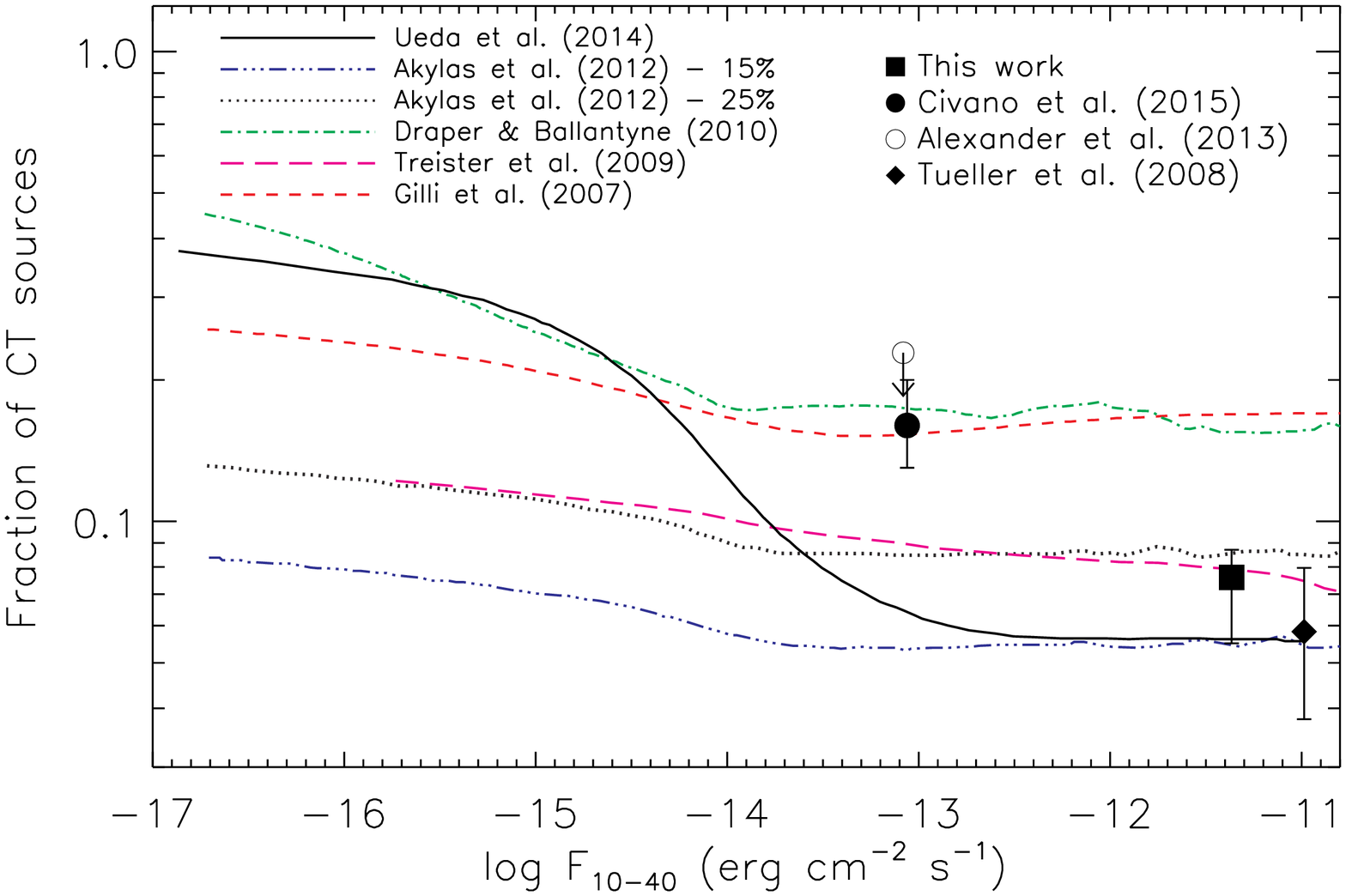}
  \caption{Fraction of CT sources predicted by different synthesis models of the CXB versus 10--40\,keV flux limit. 
 The value of $f_{\rm\,CT}$ obtained by this work is reported together with those obtained by the 9-months {\it Swift}/BAT survey (in the 14--195\,keV band; \citealp{Tueller:2008qf}) and by {\it NuSTAR} [8--24\,keV \citep{Alexander:2013ly} and 3--24\,keV band \citep{Civano:2015bh}]. The two models reported in \cite{Akylas:2012uq} consider a total fraction of CT AGN of $15\%$ and $25\%$. The plot shows that our results are in good agreement with the CXB models of  \cite{Treister:2009uq}, \cite{Akylas:2012uq} and \cite{Ueda:2014fk}.}
\label{fig:CXB_model_obsfct}
\end{figure}
\bigskip

\begin{figure}[t!]
\centering
\begin{minipage}[!b]{.48\textwidth}
\centering
\includegraphics[width=9cm]{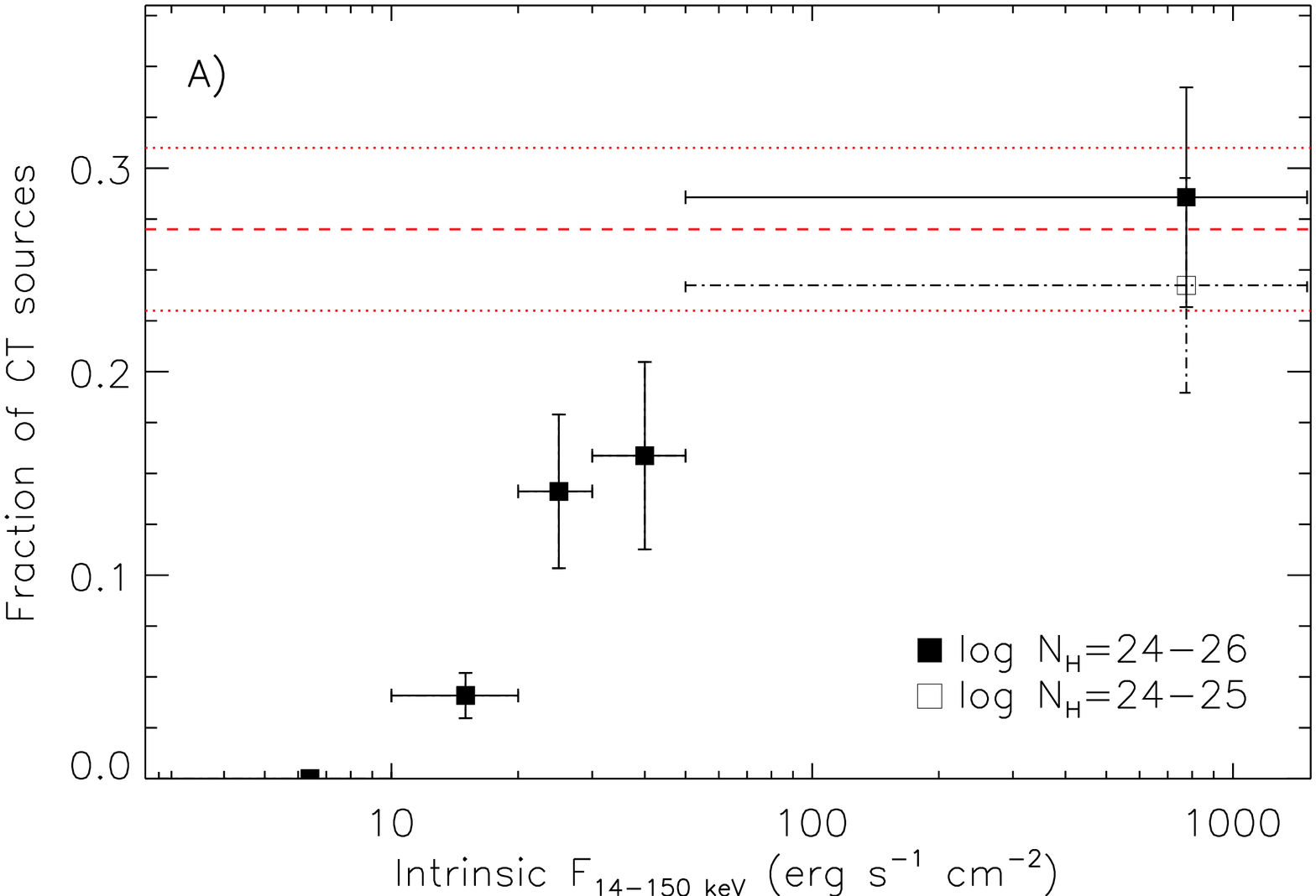}\end{minipage}
{\medskip}
\begin{minipage}[!b]{.48\textwidth}
\centering
\includegraphics[width=9cm]{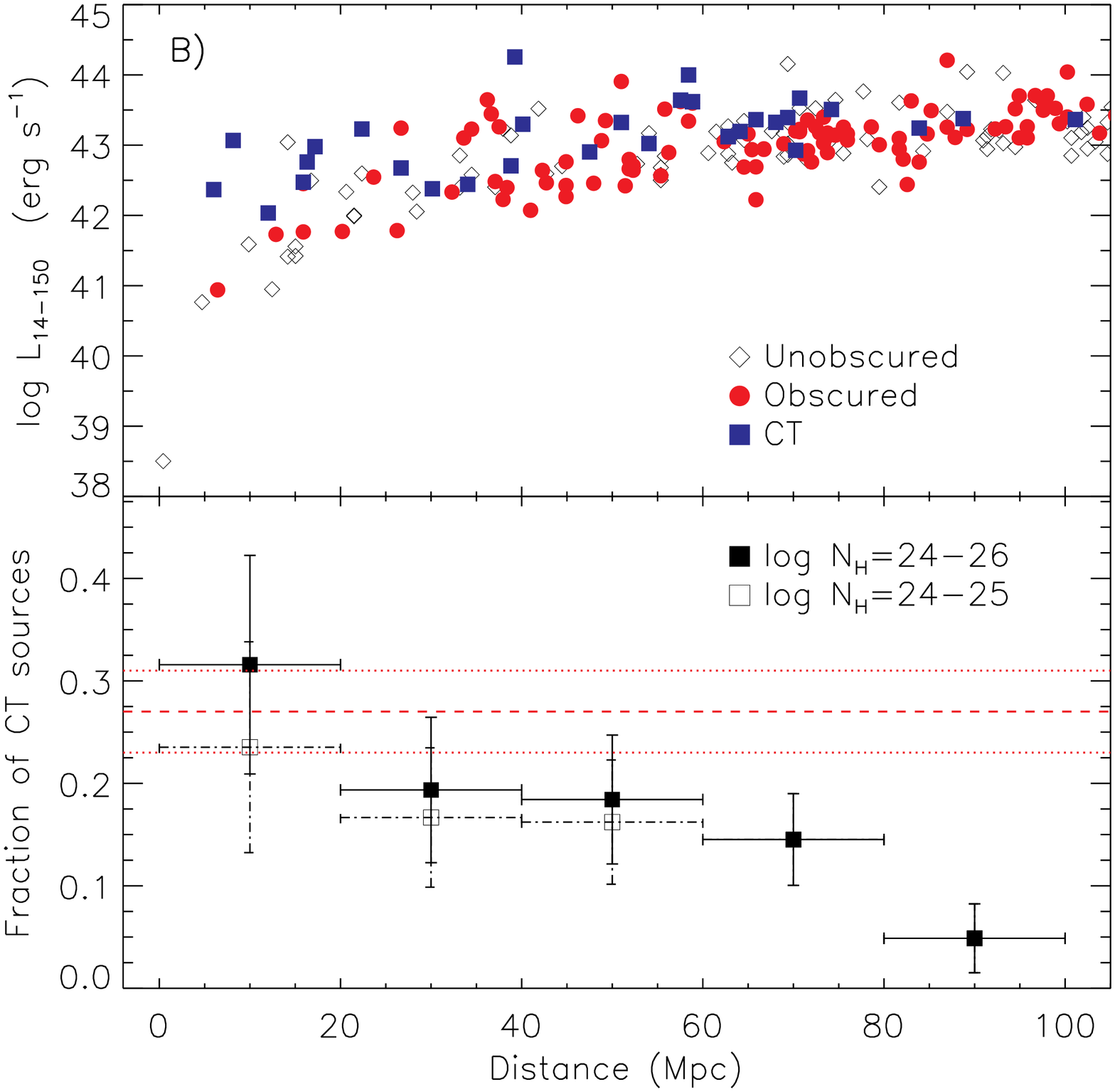}\end{minipage}
 \begin{minipage}[t]{.48\textwidth}
  \caption{{\it Panel A}: Fraction of CT AGN versus the intrinsic (i.e. absorption-corrected) 14--150\,keV flux. {\it Panel B}: Intrinsic 14--150\,keV luminosity ({\it top panel)} and observed fraction of CT sources ({\it bottom panel}) versus distance for the AGN located within 100\,Mpc. The decrease of the observed CT fraction with the decreasing intrinsic flux and the increasing distance is due to the observational bias of the {\it Swift}/BAT survey against sources with $\log N_{\rm\,H}> 24$. In both panel\,A and B the empty and filled points represent the fraction of AGN with $\log N_{\rm\,H}=24-25$ and $\log N_{\rm\,H}=25-26$, respectively. The red dashed  line shows the intrinsic fraction of sources with $\log N_{\rm\,H}=24-25$ ($27\pm4\%$, see Sect.\,\ref{sect:intrinsicNH}), while the red dotted  line is the associated uncertainty.}
\label{fig:l_z_100mpc}
 \end{minipage}
\end{figure}
\bigskip

Synthesis models of the CXB include different assumptions on the cutoff energy of the primary X-ray emission, of the intensity of the reflection component, and on the intrinsic fraction of CT sources\footnote{See for example Table\,6 of \cite{Ueda:2014fk}}. The intensity of the reprocessed X-ray emission in particular plays a rather important role in the predicted values of $f_{\rm\,CT}$ \citep{Treister:2009uq,Ricci:2011zr,Vasudevan:2013ys}. The model of \citet{Treister:2009uq} takes into account reflection from a slab using the \textsc{pexrav} model with a reflection parameter of $R = 1.2$; \cite{Akylas:2012uq} consider the same spectrum for reprocessed X-ray radiation (with $R\sim 1$), and take in account absorption by considering an X-ray source located at the centre of a uniform spherical distribution of matter.
\cite{Ueda:2014fk} consider reprocessed X-ray emission from the molecular torus by using an X-ray spectral model obtained by Monte Carlo simulations \citep{Brightman:2011oq}, taking into account also disk reflection (using \textsc{pexrav}). \cite{Ueda:2014fk} argue that assuming a strength of the Compton hump produced by the disk of $R_{\rm\,disk}=0.25$ or $R_{\rm\,disk}=1$ would change the best estimate of $f_{\rm\,CT}$ by as much as $50\%$. The significantly higher fraction of CT AGN predicted at low fluxes by the model of \cite{Ueda:2014fk} (Fig.\,\ref{fig:CXB_model_obsfct}; i.e., a factor of $\sim 3$ at $F_{10-40}=10^{-15}\rm\,erg\,cm^{-2}\,s^{-1}$ with respect to \citealp{Treister:2009uq}) is related to the increase in the fraction of obscuration at higher redshifts considered by the authors.

In Figure\,\,\ref{fig:CXB_model_obsfct} we illustrate the fraction of CT sources predicted by different synthesis models of the CXB as a function of the 10--40\,keV flux limit. The figure shows the values of $f_{\rm\,CT}$ obtained by our work, by \cite{Alexander:2013ly} and \cite{Civano:2015bh} using {\it NuSTAR}, and by \cite{Tueller:2008qf} for the 9-months {\it Swift}/BAT catalog. The {\it Swift}/BAT 70 months survey has a flux limit of $\sim 10^{-11}\rm\,erg\,s^{-1}\,cm^{-2}$ in the 14--195\,keV energy band, which corresponds to $4.3 \times 10^{-12}\rm\,erg\,s^{-1}\,cm^{-2}$ in the 10--40\,keV band (for a power-law emission with a photon index of $\Gamma=1.8$). The observed value of $f_{\rm\,CT}$ we find is in good agreement with that predicted, for the flux limit of the 70-months {\it Swift}/BAT survey, by \citet{Treister:2009uq} and \citet{Ueda:2014fk}, and with the two models of \cite{Akylas:2012uq} that consider an intrinsic fraction of CT AGN of 15\% and 25\%. In agreement with the results of \cite{Tueller:2008qf}, we find that the model of \citet{Draper:2010oq} and of \citet{Gilli:2007qf} clearly overestimate (by a factor of $\sim2$) the fraction of CT AGN for the flux-limit we probe in the 10--40\,keV band. The value of $f_{\rm\,CT}$ obtained by \citet{Civano:2015bh} combining {\it NuSTAR} detections in the 3--8, 8--24 and 3--24 keV band is larger than the value predicted by the models consistent with {\it Swift}/BAT measurements. This could be related to the large uncertainties associated with the estimation of $N_{\rm\,H}$ from hardness ratios.

\section{The intrinsic column density distribution}\label{sect:intrinsicNH}

Due to the significant effect of Compton scattering for $N_{\rm H}>10^{24}\rm\,cm^{-2}$ even hard X-ray selected samples can be biased against CT AGN (see Fig.\,\ref{fig:attenuation}). In particular only a few objects with $\log N_{\rm H} \gtrsim 24.5$ are detected by {\it Swift}/BAT. In these reflection-dominated AGN the primary X-ray emission is almost completely depleted and they are observed only through their reflection component. The effect of this observational bias is clearly illustrated in Fig.\,\ref{fig:l_z_100mpc}, which shows how the observed value of $f_{\rm\,CT}$ inferred from our sample changes with the intrinsic flux and the distance. Within 20\,Mpc the fraction of CT AGN is $f_{\rm\,CT}=32\pm11\%$, while this value clearly decreases with increasing distance, and is below $10\%$ at $D \gtrsim 80$\,Mpc.

\begin{figure}[t!]
\centering
\includegraphics[width=8.75cm]{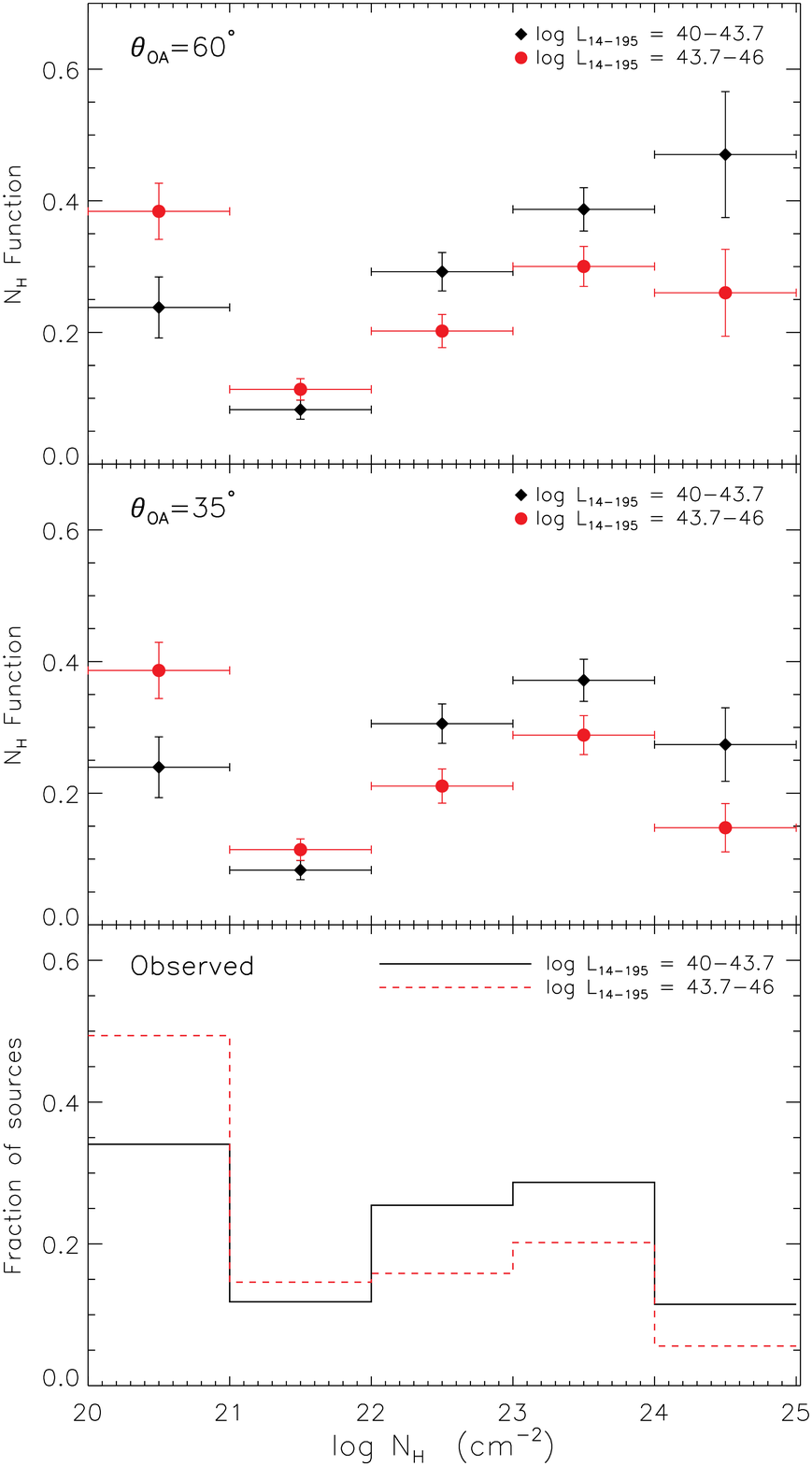}
  \caption{{\it Top panel}: $N_{\rm H}$ function of our hard X-ray selected AGN sample (normalised to unity in the $\log N_{\rm H}$ = 20--24 range) for a torus with an half-opening angle of $\theta_{\rm\,OA}=60^{\circ}$. {\it Central panel}: same as top panel for a torus with an half-opening angle of $\theta_{\rm\,OA}=35^{\circ}$. {\it Bottom panel}: observed column density distribution (normalised to unity in the $\log N_{\rm H}$ = 20--24 range). The plots show the values for 14--195\,keV luminosities in the $\log L_{\rm\,14-195} = 40-43.7$ and $\log L_{\rm\,14-195} = 43.7-46$ range. Objects with little or no absorption were assigned $\log N_{\rm H}=20$.}
\label{fig:hist_NH}
\end{figure}

To derive the intrinsic $N_{\rm H}$ distribution of AGN one must carefully correct the observed distribution (bottom panel of Fig.\,\ref{fig:hist_NH}) for selection biases. Besides absorption, the reflection components from the torus and accretion disk also affect the selection efficiency of AGN at hard X-rays. We follow the same analysis as in \citet{Ueda:2003nx,Ueda:2014fk} to constrain the ``$N_{\rm H}$ function'' [$f(L_{\mathrm{X}i},z_{i};N_{\mathrm{H}i}$)], which represents the probability distribution function of line-of-sight column density of an AGN. Since it is extremely difficult to constrain the number of AGN with $\log N_{\rm\,H}> 25$, the $N_{\rm H}$ function is normalized to unity in the range between $\log N_{\rm H}$ = 20--24 (completely unabsorbed AGN are arbitrarily assigned $\log N_{\rm H}$ = 20.0). Thanks to the large sample size, we do not assume any functional shape for the $N_{\rm H}$ function, but treat the values in discrete bins of $\log N_{\rm H}$ as independent, free parameters.  We perform a maximum-likelihood fit of the absorption function using the likelihood estimator defined by \citet{Ueda:2014fk}:
\begin{equation}
L'= -2 \sum \ln \frac{f(L_{\mathrm{X}i},z_{i};N_{\mathrm{H}i}) A(N_{\mathrm{H}i}, \Gamma_i, L_{\mathrm{X}i}, z_{i})}{\int f(L_{\mathrm{X}i},z_{i};N_{\mathrm{H}i}) A(N_{\mathrm{H}i}, \Gamma_i, L_{\mathrm{X}i}, z_{i})d\log N_{\rm\,H}}
\end{equation}
where $i$ represents each object of the sample, and $A$ is the survey area (from \citealp{Baumgartner:2013uq}) per count rate expected from a source with column density $N_{\rm\,H}$, intrinsic 14--195\,keV luminosity $L_{\rm\,X}=L_{\rm\,14-195}$ and redshift $z$. Since we are interested in the local Universe we limited the redshift range to $z<0.3$. The values of $N_{\rm H}$ obtained by our broad-band spectroscopical work have typically small uncertainties, therefore we only refer to the best-fit value of $N_{\rm H}$ of each AGN. Due to the difficulty of well constraining the column density of reflection-dominated AGN, we excluded from the sample the four most heavily Compton-thick AGN, which are likely to have $\log N_{\rm H} >25$. We obtained the count rate through the luminosity distance and the detector response by assuming the template X-ray spectra of \citet{Ueda:2014fk}, which considers both disk and torus reflection and assumes a cutoff energy of $E_{\rm\,C}=300$\,keV. The latter is taken into account using the model of \citet{Brightman:2011oq}, fixing the torus opening angle at $60^\circ$.  

The $N_{\rm H}$ function obtained is shown in the top panel of Figure\,\,\ref{fig:hist_NH} for two luminosity bins.
The number ratio between CT AGN with $\log N_{\rm H} = 24-25$ and absorbed Compton-thin AGN ($\log N_{\rm H} = 22-24$) is estimated to be $62\pm10\%$ from the entire sample. We find a decrease in the fraction of obscured Compton-thin AGN for increasing luminosities as found by several previous studies (e.g., \citealp{Beckmann:2009fk}), with the $N_{\rm\,H}$ function going from $68\pm4\%$ for the low-luminosity bin ($\log L_{\rm\,14-195} = 40-43.7$) to $50\pm4\%$ for the high-luminosity one ($\log L_{\rm\,14-195} = 43.7-46$). The $N_{\rm\,H}$ function at $\log N_{\rm H} = 24-25$ has also a significantly higher value in the low-luminosity bin ($f=47\pm10\%$) than in the high-luminosity one ($f=30\pm3\%$). Normalizing the $N_{\rm\,H}$ distribution to unity in the $\log N_{\rm H} = 20-25$ range we obtain that the fraction of obscured Compton-thin sources decreases from $46\pm3\%$ ($\log L_{\rm\,14-195} = 40-43.7$) to $39\pm3\%$ ($\log L_{\rm\,14-195} = 43.7-46$), while the fraction of CT AGN with $\log N_{\rm\,H}=24-25$ is $32\pm7\%$ and $21\pm5\%$ for the low and the high luminosity bin, respectively. The fraction of CT AGN with $\log N_{\rm\,H}=24-25$ for the whole sample is $27\pm4\%$. This value is larger than that predicted at low $z$ by \cite{Aird:2015vn}, but is in very good agreement with that inferred by \cite{Burlon:2011uq} ($20^{+9}_{-6}\%$) using a smaller sample of $\sim 200$ {\it Swift}/BAT selected AGN.

The correction factors we calculated to reproduce the intrinsic number of CT AGN depend on the geometry of the absorbing and reprocessing material. To illustrate this effect we show in the central panel of Fig.\,\ref{fig:hist_NH} the $N_{\rm H}$ function obtained assuming a torus opening angle of $\theta_{\rm\,OA}=35^\circ$. 
While the population of Compton-thin AGN is unchanged, adopting this geometry we found that the number ratio between CT AGN with $\log N_{\rm H} = 24-25$ and absorbed Compton-thin AGN ($\log N_{\rm H} = 22-24$) is $36\pm5\%$ for the whole sample, lower than the value obtained for $\theta_{\rm\,OA}=60^\circ$. Similarly to what we obtained for $\theta_{\rm\,OA}=60^\circ$, we found that the $N_{\rm\,H}$ function at $\log N_{\rm H} = 24-25$ has a significantly higher value in the low-luminosity bin ($f=27\pm6\%$) than in the high-luminosity one ($f=15\pm4\%$). Normalizing the $N_{\rm\,H}$ distribution to unity in the $\log N_{\rm H} = 20-25$ range we found that the fraction of CT AGN with $\log N_{\rm\,H}=24-25$ is $22\pm4\%$ and $13\pm3\%$ for the low and high-luminosity bin, respectively, and $17\pm3\%$ for the whole sample.

The fraction of obscured Compton-thin AGN of our sample significantly decreases with increasing luminosity. This trend has been explained with the decrease of the covering factor of the circumnuclear material with the luminosity, possibly due to radiation pressure (e.g., \citealp{Lusso:2013cr} and references therein). We also find significant evidence for a luminosity-dependence of the fraction of CT sources, both assuming $\theta_{\rm\,OA}=60^\circ$ and $\theta_{\rm\,OA}=35^\circ$. Studying {\it NuSTAR} observations of 10 AGN, \citet{Brightman:2015cr} recently found possible evidence of a strong decrease of the covering factor of the torus with the luminosity ($f_{\rm\,CT}\propto -0.41 \log L_{2-10}$), similarly to what was found for Compton-thin objects. At higher redshifts \cite{Buchner:2015ve} found instead that the fraction of CT AGN is compatible with being constant with the luminosity.

\section{Summary and conclusions}
In this work we report the first results obtained by the largest study of hard X-ray selected AGN to date, which includes the 834 AGN of the 70-months {\it Swift}/BAT catalog. We find 55\,CT AGN, which represents $7.6^{+1.1}_{-2.1}\%$ of the total population of non-blazar objects. We find the first evidence of CT obscuration in 26 objects, increasing considerably the number of hard X-ray selected CT AGN at low redshift. The observed fraction of CT AGN we infer for the flux limit of the {\it Swift}/BAT survey is in agreement with the recent CXB synthesis models of \citet{Treister:2009uq}, \cite{Akylas:2012uq} and \citet{Ueda:2014fk}.

We reconstruct the intrinsic column density distribution of AGN and find that the fraction of obscured Compton-thin AGN ($\log N_{\rm H} = 22-24$) varies between $\sim46\%$ (for $\log L_{\rm\,14-195} = 40-43.7$) and $\sim39\%$ ($\log L_{\rm\,14-195} = 43.7-46$). We also find a decrease in the fraction of CT AGN for increasing luminosity. We show that the intrinsic fraction of CT AGN with $\log N_{\rm\,H}=24-25$ normalised to unity in the $\log N_{\rm H} = 20-25$ range is $27\pm4\%$. This is consistent with the observed value obtained for AGN located within 20\,Mpc (see Fig.\,\ref{fig:l_z_100mpc}), which shows that a significant fraction of AGN in the local Universe are heavily obscured.

Future {\it NuSTAR} observations of nearby AGN, selected for example by their [OIII] emission, will allow to detect the CT AGN missing from the {\it Swift}/BAT survey. At the same time deep {\it NuSTAR} observations are expected to resolve $\sim 30\%$ of the integrated flux of the CXB \citep{Ballantyne:2011nx}, and will shed light on the fraction of CT sources at lower flux levels, which will allow to even better discriminate between different synthesis models of the CXB.

\acknowledgments
We thank the referee for the very detailed and insightful report, which helped us to improve the quality of the manuscript. CR and FEB acknowledge financial support from the CONICYT-Chile ''EMBIGGEN" Anillo (grant ACT1101). Part of this work was carried out while CR was Fellow of the Japan Society for the Promotion of Science (JSPS). We thank L. Burtscher, J. Buchner, C. S. Chang and C. Vignali for their comments on the manuscript. We acknowledge financial support from FONDECYT 1141218 (CR, FEB), Basal-CATA PFB--06/2007 (CR, FEB), and the Ministry of Economy, Development, and Tourism's Millennium Science Initiative through grant IC120009, awarded to The Millennium Institute of Astrophysics, MAS (FEB). This work was partly supported by the Grant-in-Aid for Scientific Research 26400228 (YU) from the Ministry of Education, Culture, Sports, Science and Technology of Japan (MEXT). MK acknowledges support from the Swiss National Science Foundation (SNSF) through the Ambizione fellowship grant PZ00P2\textunderscore154799/1.

{\it Facilities:} \facility{Swift}, \facility{XMM-Newton}, \facility{Suzaku}, \facility{ASCA}, \facility{Chandra}.


%
\appendix
%
The list of CT sources, together with their characteristics inferred by our broad-band X-ray spectroscopical study are reported in Table\,\ref{tab:CTlist}.

\clearpage

\begin{landscape}
\LongTables
\begin{deluxetable*}{lllccrrrccl}
\tabletypesize{\footnotesize}
\tablecaption{CT AGN in the {\it Swift}/BAT 70-months catalog.\label{tab:CTlist}}
\tablewidth{0pt}
\tablehead{
\colhead{(1) } &\colhead{(2)} &\colhead{(3)} & \colhead{(4) } & \colhead{(5)}  & \colhead{(6)} & \colhead{(7)} &  \colhead{(8) } &  \colhead{ (9) }  &  \colhead{ (10) } &  \colhead{ (11) } \\
\noalign{\smallskip}
\colhead{SWIFT ID} & \colhead{Counterpart} & \colhead{Type} & \colhead{z }  & \colhead{ $\log N_{\rm H}$} & \colhead{Fe K$\alpha$ EW} &  \colhead{$\Gamma_{2-10}$}&  \colhead{$\Gamma$}  &    \colhead{$\log L_{\rm\,2-10}$} &\colhead{$\log L_{\rm\,14-150}$} &\colhead{Facility} \\
\noalign{\smallskip}
\colhead{ } &\colhead{ } &\colhead{ } &\colhead{ } & \colhead{ [\scriptsize{$\rm\,cm^{-2}$}]} & \colhead{ [\scriptsize{eV}]}  & \colhead{ } & \colhead{ } &  \colhead{[\scriptsize{$\rm\,erg\,s^{-1}$}] } &  \colhead{[\scriptsize{$\rm\,erg\,s^{-1}$}] }  &\colhead{ } 
}
\startdata

{\bf SWIFT\,J0030.0$-$5904} 			&	ESO 112$-$6				    	&  	\nodata	&  	0.0290     &	  24.03 [23.79 -- 24.43]	  &   $7300^{+4056}_{-5641}$     	&   $-0.70^{+5.01}_{-NC}$			& $1.78^{+0.35}_{-0.38}$	&	       43.12	&	   43.53   & SX	  			\\  
{\bf SWIFT\,J0105.5$-$4213} 			&	MCG$-$07$-$03$-$007         	&   2      	&  	0.0302     &      24.18 [23.95 -- 24.30]	  &   $\leq 2246$               	&	$-1.30^{+1.29}_{-1.31}$			& $2.12^{+0.25}_{-0.41}$	&	       43.46	&	   43.55   & SX	   \\    
SWIFT\,J0111.4$-$3808 					&	NGC 424                         &	1.9     &  	0.0118     &      24.33 [24.32 -- 24.34]	  &   $928_{-72}^{+109}$        	&	$0.11^{+0.06}_{-0.06}$			& $2.64^{+0.11}_{-0.39}$ 	&	       43.77	&	   43.32   & XE	   \\    
{\bf SWIFT\,J0122.8$+$5003}$^{\rm B}$ 	&	MCG +08$-$03$-$018              &  	2      	&   0.0204     &      24.24 [24.09 -- 24.58]	  &   $\leq 930$               		&	$-0.29^{+0.36}_{-0.97}$			& $2.71^{+0.05}_{-0.55}$	&	       43.98	&	   43.38   & SX 		      \\ 
{\bf SWIFT\,J0128.9$-$6039} 			&	2MASX\,J01290761$-$6038423 	    &   \nodata	&   0.2030     &      24.13 [23.90 -- 24.32]	  &   $\leq 1204$               	&	$-0.27^{+1.00}_{-1.63}$			& $2.18^{+0.21}_{-0.36}$	&	       45.23	&	   45.23   & SX		     \\  
{\bf SWIFT\,J0130.0$-$4218}$^{\dagger}$ &	ESO 244$-$IG 030                &	2      	&  	0.0256     &      24.20 [24.02 -- 24.55]	  &   NC               				&	$0.16^{+1.03}_{-2.60}$			& $2.45^{+0.35}_{-0.40}$	&	       43.68	&	   43.42   & SX	   \\    
SWIFT\,J0242.6$+$0000 					&	NGC 1068                        &   2      	& 	0.0038     &      24.95 [24.63 -- 25.16]	  &   $565^{+58}_{-20}$        		&	$0.89^{+0.06}_{-0.06}$			& $2.37^{+0.10}_{-0.08}$	&	       42.93	&	   42.76   & XE	     \\  
{\bf SWIFT\,J0250.7$+$4142} 			&	NGC 1106                        &	2$^{1}$	& 	0.0145     &      24.25 [24.08 -- 24.54]	  &   $\leq 2558$               	&	$0.83^{+1.21}_{-2.28}$			& $1.87^{+0.26}_{-0.29}$	&	       42.80	&	   43.12   & SX	     \\  
SWIFT\,J0251.3$+$5441$^{\rm A}$ 		&	2MFGC 02280                     &  	2      	& 	0.0152     &      24.06 [23.96 -- 24.18]	  &   $336_{-331}^{+748}$        	&	$-2.17^{+0.67}_{-NC}$			& $1.86^{+0.13}_{-0.21}$	&	       43.02	&	   43.36   & SX       \\
{\bf SWIFT\,J0251.6$-$1639} 			&	NGC 1125                        &   2      	& 	0.0110     &      24.27 [24.03 -- 24.59]	  &   $\leq 2023$               	&	$0.03^{+1.47}_{-1.93}$			& $2.01^{+0.29}_{-0.18}$	&	       42.74	&	   42.90   & SX	       \\
SWIFT\,J0304.1-0108						&	NGC 1194						&  1.9$^{1}$&	0.0136	   &	  24.33 [24.30 -- 24.39]	  &	  $768^{+235}_{-241}$			& 	$0.21^{+0.12}_{-0.13}$	 		& $2.00^{+0.12}_{-0.08}$	& 	       43.69	&	   43.62   & XE		\\		
{\bf SWIFT\,J0308.2$-$2258} 			&	NGC 1229                        &	2      	& 	0.0360     &      24.94 [24.49 -- NC]\phantom{a}\,\,		  &   $\leq 1662$               	&	$-1.51^{+1.20}_{-1.36}$			& $2.14^{+0.43}_{-0.24}$	&	       43.96	&	   44.10   & SX	     \\  
SWIFT\,J0350.1$-$5019 					&	ESO 201$-$4				    	&  	2      	&   0.0359     &      24.32 [24.25 -- 24.35]	  &   $503^{+136}_{-106}$        	&	$-0.40^{+0.25}_{-0.26}$			& $2.09^{+0.16}_{-0.22}$	&	       44.44	&	   44.28   & XE	     \\  
{\bf SWIFT\,J0357.5$-$6255} 			&	2MASX\,J03561995$-$6251391      &   1.9     &   0.1076     &      24.17 [23.99 -- 24.38]	  &   $\leq 542$        			&	$-2.77^{+2.08}_{-NC}$			& $2.26^{+0.27}_{-0.32}$	&	       44.84	&	   44.77   & SX	   \\  
{\bf SWIFT\,J0427.6$-$1201}$^{\dagger}$ &	MCG$-$02$-$12$-$017 			&	2      	& 	0.0325     &      24.25 [23.79 -- 25.26]	  &   NC        					&	$-0.59^{+1.24}_{-NC}$			& $1.94^{+0.38}_{-0.43}$	&	       43.41	&	   43.67   & SX	     \\
SWIFT\,J0453.4$+$0404 					&	CGCG 420$-$015                  &  	2      	&   0.0294     &      24.14 [23.93 -- 24.18]	  &   $450^{+1725}_{-24}$        	&	$-0.46^{+0.25}_{-0.26}$			& $2.27^{+0.12}_{-0.40}$	&	       44.00	&	   43.93   & XE     \\   
SWIFT\,J0601.9$-$8636 					&	ESO 005$-$ G 004                &   2      	& 	0.0062     &      24.34 [24.28 -- 24.44]	  &   $1414^{+1175}_{-1242}$     	&   $-0.86^{+0.22}_{-0.22}$			& $1.81^{+0.14}_{-0.11}$	&	       42.78	&	   42.68   & SuX		     \\  
SWIFT\,J0615.8$+$7101 					&	Mrk 3                           &	1.9     & 	0.0135     &      24.07 [24.03 -- 24.13]	  &   $354_{-9}^{+32}$        		&	$-0.52^{+0.04}_{-0.04}$			& $1.92^{+0.10}_{-0.05}$	&	       43.67	&	   44.00   & XE    \\    
{\bf SWIFT\,J0656.4$-$4921} 			&	2MASX\,J06561197$-$4919499      &  	2      	&   0.0410     &      24.03 [23.93 -- 24.33]	  &   $\leq 2224$               	&	$-2.03^{+1.77}_{-0.81}$			& $1.91^{+0.31}_{-0.25}$	&	       43.48	&	   43.77   & SX	   \\    
{\bf SWIFT\,J0714.2$+$3518}$^{\rm A}$  			&	MCG +06$-$16$-$028    &   1.9     &   0.0157     &      24.80 [24.05 -- NC]\phantom{a}\,\,		  &   $\leq 6136$               	&	$-1.65^{+1.77}_{-1.12}$			& $1.92^{+0.42}_{-0.59}$	&	       43.04	&	   43.33   & SX	   \\    
{\bf SWIFT\,J0743.0$+$6513}$^{\rm B}$ 	&	Mrk 78                          &	2      	& 	0.0371     &      24.11 [23.99 -- 24.19]	  &   $366^{+2083}_{-110}$       	& 	$-0.49^{+0.51}_{-0.53}$			& $2.49^{+0.21}_{-0.36}$	&	       43.82	&	   43.53   & XE      \\  
SWIFT\,J0807.9$+$3859 					&	Mrk 622                         &  	1.9     & 	0.0232     &      24.29 [23.99 -- NC]\phantom{a}\,\,		  &   $597^{+567}_{-324}$        	&	$-1.09^{+0.67}_{-0.74}$			& $2.10^{+0.22}_{-0.84}$	&	       43.26	&	   43.37   & XE     \\   
{\bf SWIFT\,J0902.7$-$6816}$^{\rm B}$ 	&	NGC 2788A            		    &	2      	& 	0.0133     &      25.55 [24.14 -- NC]\phantom{a}\,\,		  &   $ 3408^{+1770}_{-3232}$    	&   $-1.74^{+1.36}_{-NC}$			& $1.58^{+0.68}_{-0.11}$	&	       43.03	&	   43.64   & SX     \\  
{\bf SWIFT\,J0919.2$+$5528} 			&	Mrk 106				     	    &  	1.9     &  	0.1234     &  	  24.01 [23.86 -- 24.15]	  &   $\leq 1736$               	&	$-1.31^{+0.98}_{-1.01}$			& $2.12^{+0.41}_{-0.17}$	&	       44.54	&	   44.62   & SX	    \\   
{\bf SWIFT\,J0924.2$-$3141} 			&	2MASX\,J09235371$-$3141305      &   2      	&   0.0424     &      24.11 [24.03 -- 24.20]	  &   $\leq 1683$        			&	$-2.10^{+0.99}_{-NC}$			& $2.19^{+0.24}_{-0.11}$	&	       44.14	&	   44.04   & SX	     \\
SWIFT\,J0934.7$-$2156        			& 	ESO 565$-$G019            		&   2$^{1}$	&  0.0163	   &	  24.65 [24.48 -- NC]\phantom{a}\,\,		  &	  $1100^{+600}_{-600}$			&	$0.60^{+0.52}_{-0.38}$			& $1.86^{+0.21}_{-0.37}$	&	       43.50	&	   43.67   & SuX 					\\		
SWIFT\,J0935.9$+$6120 					&	MCG +10$-$14$-$025              &  	1.9     &   0.0394     &      24.35 [24.31 -- 24.45]	  &   $\geq 222$ 	       			&	$0.01^{+0.34}_{-0.36}$			& $2.33^{+0.25}_{-0.25}$	&	       44.23	&	   44.07   & XE    \\  
SWIFT\,J1001.7$+$5543$^{\rm A}$			&	NGC 3079                        &   1.9     & 	0.0037     &      25.10 [24.51 -- NC]\phantom{a}\,\,		  &   $1609^{+389}_{-1541}$      	&  	$0.90^{+0.50}_{-0.47}$			& $1.46^{+0.09}_{-0.08}$	&	       41.30	&	   42.47   & XE	    \\   
{\bf SWIFT\,J1031.5$-$4205}$^{\dagger}$ &	ESO 317$-$ G 041  				&	\nodata	& 	0.0193     &      24.30 [24.08 -- 24.73]	  &   NC         					&	$-2.98^{+0.56}_{-NC}$			& $2.20^{+0.27}_{-0.22}$	&	       43.27	&	   43.24   & SX	     \\  
{\bf SWIFT\,J1033.8$+$5257} 			&	SDSS\,J103315.71+525217.8	    &  	\nodata	& 	0.0653     &      24.27 [24.06 -- 24.53]	  &   $\leq 1776$               	&	$-0.93^{+0.55}_{-1.62}$			& $2.35^{+0.24}_{-0.30}$		&	       44.33	&	   44.18   & SX	     \\  
SWIFT\,J1048.4$-$2511$^{\rm A}$ 					&	NGC 3393              &   2      	& 	0.0125     &      24.50 [24.31 -- 24.82]	  &   $1585^{+467}_{-466}$       	& 	$0.70^{+0.56}_{-0.55}$			& $1.79^{+0.18}_{-0.22}$	&	       42.63	&	   43.02   & XE	    \\   
SWIFT\,J1206.2$+$5243 					&	NGC 4102                   	    &	2      	& 	0.0028     &      24.18 [24.06 -- 24.27]	  &   $275 ^{+234 }_{-204 }$     	&   $0.90^{+0.47}_{-0.45}$			& $1.80^{+0.18}_{-0.23}$	&	       41.66	&	   42.04   & XE     \\  
{\bf SWIFT\,J1212.9$+$0702}$^{\dagger}$ &	NGC 4180                   	    &  	2$^{1}$	& 	0.0070     &      24.15 [23.93 -- 24.42]	  &   NC               				&	$-2.78^{+2.62}_{-NC}$			& $1.75^{+0.29}_{-0.19}$	&	       41.92	&	   42.38   & SX	     \\  
SWIFT\,J1253.5$-$4137 					&	ESO 323$-$32                    &   2      	&   0.0160     &      24.79 [24.39 -- NC]\phantom{a}\,\,		  &   $1787^{+226}_{-217}$       	& 	$-0.01^{+0.24}_{-0.25}$			& $1.96^{+0.42}_{-0.58}$	&	       43.16	&	   43.39   & SuX		    \\   
SWIFT\,J1305.4$-$4928 					&	NGC 4945                        &	2$^{1}$	& 	0.0019     &      24.80 [24.76 -- 24.93]	  &   $863^{+46}_{-42}$        		&	$-0.04^{+0.06}_{-0.06}$			& $1.80^{+0.05}_{-0.06}$	&	       42.07	&	   43.07   & XE	      \\ 
SWIFT\,J1412.9$-$6522 					&	Circinus Galaxy            	    &   2$^{1}$	& 	0.0014     &      24.40 [24.39 -- 24.41]	  &   $2019^{+224}_{-21}$        	&	$0.21^{+0.02}_{-0.02}$			& $2.50^{+0.01}_{-0.01}$	&	       42.63	&	   42.37   & XE	     \\  
SWIFT\,J1416.9$-$4640 					&	IGR\,J14175$-$4641              &	2      	&   0.0766     &      24.35 [24.20 -- 24.54]	  &   $\leq 938$               		&	$-0.30^{+0.94}_{-1.38}$			& $2.11^{+0.17}_{-0.24}$	&	       44.71	&	   44.80   & SX			   \\    
SWIFT\,J1432.8$-$4412 					&	NGC 5643                        &  	2      	& 	0.0040     &      25.40 [25.06 -- NC]\phantom{a}\,\,		  &   $1640_{-13}^{+2484}$       	& 	$-0.30^{+0.12}_{-0.12}$			& $1.65^{+0.13}_{-0.11}$	&	       42.43	&	   42.98   & XE	   \\    
SWIFT\,J1442.5$-$1715 					&	NGC 5728                        &   2      	& 	0.0093     &      24.13 [24.09 -- 24.16]	  &   $780^{+1561}_{-105}$       	& 	$-1.20^{+0.17}_{-0.16}$			& $1.95^{+0.03}_{-0.04}$	&	       42.86	&	   43.30   & SuX		     \\  
{\bf SWIFTJ1445.6+2702}$^{\rm A}$    		&	CGCG\,164$-$019					&	1.9$^{1}$ &	0.0299	   &	  24.75 [24.60 -- 25.49]	  &			$\leq 1282$				&	$0.10^{+0.70}_{-0.81}$			& $2.15^{+0.32}_{-0.21}$	&	       44.57	&	   44.65   & SX					 \\	
SWIFT\,J1635.0$-$5804 					&	ESO 137$-$ G 034                &	2      	& 	0.0090     &      24.30 [24.23 -- 24.37]	  &   $1112^{+2178}_{-98}$       	& 	$0.02^{+0.23}_{-0.23}$			& $2.14^{+0.17}_{-0.17}$	&	       42.65	&	   42.71   & XE      \\  
{\bf SWIFT\,J1643.2$+$7036}$^{\rm A}$ 	&	NGC 6232          			   	&  	2      	& 	0.0148     &      24.94 [24.53 -- NC]\phantom{a}\,\,		  &  $\leq 3314$         			&	$-0.39^{+1.76}_{-2.16}$			& $1.80^{+0.83}_{-0.35}$	&	       42.82	&	   43.19   & SX     \\  
SWIFT\,J1652.0$-$5915A					& 	ESO 138$-$ G 001                &   2$^{1}$	& 	0.0091     &      25.25 [24.94 -- NC]\phantom{a}\,\,		  &    $885^{+79}_{-84}$        	& 	$-0.06^{+0.09}_{-0.09}$			& $2.19^{+0.07}_{-0.07}$	&	       44.09	&	   44.26   & XE	    \\   
SWIFT\,J1652.9$+$0223 					&	NGC 6240                        &	1.9     & 	0.0245     &      24.40 [24.32 -- 24.45]	  &   $357^{+588}_{-52}$        	&	$0.88^{+0.16}_{-0.16}$			& $2.33^{+0.06}_{-0.10}$	&	       44.75	&	   44.62   & XE	     \\  
SWIFT\,J1800.3$+$6637 					&	NGC 6552                        &  	2$^{1}$	& 	0.0265     &      24.05 [23.83 -- 24.40]	  &   $756^{+411}_{-522}$        	&	$-2.66^{+1.75}_{-NC}$			& $1.87^{+0.28}_{-0.30}$	&	       43.27	&	   43.59   & XE	     \\  
{\bf SWIFT\,J2015.2$+$2526}$^{\dagger}$ &	2MASX\,J20145928+2523010        &   2      	& 	0.0453     &      24.42 [24.25 -- 24.62]	  &   NC               		    	&   $1.29^{+1.19}_{-1.13}$			& $2.30^{+0.25}_{-0.29}$	&	       44.59	&	   44.50   & SX	       \\
{\bf SWIFTJ2028.5+2543}$^{\rm B,C}$		&	NGC 6921					 	&  	\nodata	& 	0.0145	   &	  24.27 [24.09 -- 24.39]	  &   $944^{+373}_{-292}$			&	$0.47_{-0.34}^{+0.33}$			& $1.96^{+0.14}_{-0.12}$	&	       42.88 	& 	   43.12   & XE	  \\				
{\bf SWIFT\,J2102.6$-$2810} 			&	ESO 464$-$ G016  				&  	2      	&   0.0364     &      24.19 [23.94 -- 24.59]	  &  $\leq 909$               		&	$-1.86^{+1.82}_{-NC}$			& $2.00^{+0.24}_{-0.32}$	&	       43.74	&	   43.91   & SX	     \\  
SWIFT\,J2148.3$-$3454 					&	NGC 7130                        &   1.9     & 	0.0162     &      24.00 [23.90 -- 24.21]	  &   $\leq 1716$        			&	$1.07^{+0.73}_{-0.62}$			& $2.05^{+0.09}_{-0.08}$	&	       42.13	&	   42.93   & CA		     \\  
SWIFT\,J2207.3$+$1013$^{\rm A}$ 		&	NGC 7212 NED02                  &	2      	& 	0.0267     &      24.41 [24.34 -- 24.48]	  &   $\geq 454$        			&	$-0.12^{+0.22}_{-0.22}$			& $2.20^{+0.32}_{-0.39}$	&	       44.41	&	   43.94   & XE     \\  
{\bf SWIFT\,J2242.4$-$3711}$^{\dagger}$	&	ESO 406$-$ G 004            	&  	\nodata	& 	0.0290     &      24.74 [24.19 -- NC]\phantom{a}\,\,		  &   NC               				&	\multicolumn{1}{c}{NC}			& $2.04^{+0.78}_{-0.96}$	&	       43.44	&	   43.59   & SX		       \\
SWIFT\,J2304.9$+$1220 					&	NGC 7479                        &   1.9     & 	0.0079     &      24.16 [24.03 -- 24.28]	  &   $684^{+259}_{-672}$        	&	$-1.59^{+1.20}_{-1.27}$			& $1.81^{+0.26}_{-0.26}$	&	       42.07	&	   42.44   & XE	       \\
{\bf SWIFT\,J2307.9$+$2245}$^{\dagger}$ &	2MASX\,J23074887+2242367	    &	2$^{1}$	& 	0.0350     &      24.20 [24.00 -- 24.50]	  &   NC               				&	\multicolumn{1}{c}{NC}			& $1.89^{+0.28}_{-0.30}$	&	       43.48	&	   43.79   & SX	     \\  
SWIFT\,J2318.4$-$4223 					&	NGC 7582                        &	2      	&	0.0052     &      24.33 [24.32 -- 24.34]	  &   $515_{-6}^{+150}$        		&	$-0.15^{+0.04}_{-0.04}$			& $2.33^{+0.05}_{-0.06}$	&	       43.48	&	   43.23   & XE	     \\  
SWIFT\,J2328.9$+$0328 					&	NGC 7682                        &	1.9     &	0.0171     &      24.30 [24.18 -- 24.38]	  &   $\geq 63$               		&	$-0.61^{+0.44}_{-0.47}$			& $2.18^{+0.18}_{-0.21}$	&	       43.50	&	   43.51   & XE	     \\

\enddata


\tablecomments{The table reports the (1) {\it Swift} ID, (2) counterpart name, (3) optical classification, (4) redshift, (5) value and $90\%$ confidence interval of the column density, (6) Fe K$\alpha$ EW, (7) photon index obtained by fitting the spectrum in the 2--10\,keV range with a power-law, (8) photon index obtained by fitting the 0.3--150\,keV spectrum (see Sect.\,\ref{Sect:specanalysis}), (9) 2--10\,keV and (10) 14--150\,keV intrinsic (i.e. absorption and k-corrected) luminosities, and (11) X-ray observatory used for the soft X-ray spectra (CA={\it Chandra}/ACIS; SuX={\it Suzaku}/XIS; SX={\it Swift}/XRT; XE={\it XMM-Newton}/EPIC). The optical classifications are taken from the {\it Swift}/BAT Spectral Survey (BASS paper I, Koss et al. in prep.) unless stated otherwise. BASS will report the optical characteristics of more than 500 BAT selected AGN. Dots are reported when no optical classification is available. Newly identified CT AGN are reported in boldface. }
\tablenotetext{A}{Sources reported to be CT by Koss et al. (in prep.) thanks to {\it NuSTAR} observations.}
\tablenotetext{B}{Sources reported to be Compton-thin by previous works due to the narrower energy band used for the X-ray spectral analysis.}
\tablenotetext{C}{{\it Swift}/BAT flux due to the combined emission of NGC\,6921 and MCG\,+04$-$48$-$002. The 14--150\,keV luminosity reported refers only to NGC\,6921}
\tablenotetext{1}{Optical classification from the literature.}
\tablenotetext{$^{\dagger}$}{Fe K$\alpha$ line not detected because of the low signal-to-noise ratio of the {\it Swift}/XRT observation.}
\tablenotetext{NC}{value not constrained.}

\end{deluxetable*}
\clearpage
\end{landscape}

\end{document}